\documentclass[aps,prl,showpacs,twocolumn]{revtex4}
\usepackage{newlfont}
\usepackage{amssymb}
\usepackage{amsfonts}
\usepackage{amsmath}
\usepackage{color}

\newtheorem{defi}{Definition}

\newcommand{\tr}{\mbox{Tr}}

\newcommand{\ket}[1]{\mbox{$| #1 \rangle$}}
\newcommand{\bk}[2]{\ensuremath{\langle #1 | #2 \rangle}}


\begin{document}
\title{Local unitary equivalence and distinguishability of arbitrary
multipartite pure states}

\author{Adam Sawicki}
\affiliation{Center for Theoretical Physics PAS, Al. Lotnik\'ow 32/46, 02-668
Warszawa, Poland}

\author{Marek Ku\'s}
\affiliation{Center for Theoretical Physics PAS, Al. Lotnik\'ow 32/46, 02-668
Warszawa, Poland}

\date{\today}

\begin{abstract}
We give an universal algorithm for testing the local unitary equivalence of
states for multipartite system with arbitrary dimensions.
\end{abstract}

\maketitle

Entanglement in multipartite quantum systems, due to its importance in
quantum information theory, became a topic of numerous investigations (see
\cite{amico08} for a recent comprehensive review of various aspects of
characterization and applications of multipartite entangled states). Until
now, however, the efforts did not bring final solutions to many fundamental
problems. One of the question is the classification of states which are
interconvertible by local unitary transformations, i.e.\ operations on the
whole systems composed from unitary actions (purely quantum evolutions) each
of which is restricted to a single subsystem. To appreciate the importance of
such an experimental setting let us remaind that it is a basis for such
spectacular applications of quantum information technologies like
teleportation or dense codding where the fundamental parts of experiments
consist of manipulations restricted to parts of the whole systems in distant
laboratories.

Recently, the problem of checking local unitary (LU) equivalence of qubit
systems was solved in \cite{kraus10,kraus10a}. In this letter we present a
geometric approach to the problem which supersedes previous ones by its
applicability to systems of arbitrary dimensions treating on equal footing
qubits and qudits. The same geometric formulations allows also to answer in
arbitrary dimensions the question of distinguishing multipartite states by
local measurements (i.e.\ measurements constrained to single subsystems). By
taking advantage of relations between both concepts we are able to give an
universal algorithm for checking LU-equivalence of pure states in arbitrary
dimensions. Despite a rather abstract geometrical picture underlying our
approach the final algorithms for checking equivalence and distinguishability
are easy in applications.

Consider a quantum system consisting of $M$ identical subsystems with the
Hilbert space $\mathcal{H}=\otimes _{i=1}^M\mathcal{H}_i$ where each
$\mathcal{H}_i$ is $N$-dimensional
dimensions of the subsystems can be done along the same lines, albeit with
. We say that two pure states $\ket{v_1}\in \mathcal{H}$ and
$\ket{v_2}\in \mathcal{H}$ are locally unitary equivalent (LU-equivalent) if
and only if there exist unitary operators $U_i\in \mathrm{U}(N)$ such that
\begin{equation}\label{LUequiv}
\ket{v_1}=U_1\otimes\ldots\otimes U_M \ket{v_2}.
\end{equation}
We call states $\ket{v_1}$ and $\ket{v_2}$ indistinguishable by
local measurements if and only if
\begin{equation}\label{indisti}
\bk{v_1}{Av_1}=\bk{v_2}{Av_2}\quad \forall A,
\end{equation}
where $A$ is a local hermitian operator, i.e, it is real combinations of
$A_1\otimes I\otimes\ldots\otimes I,\,\ldots\,,\,I\otimes \ldots\otimes
I\otimes A_M$ and $A_i$ are $N\times N$ hermitian matrices. Each operator
correspond to a measurement performed on a single subsystem. Physically, we
may imagine that subsystems are located in different laboratories, each local
measurement is performed on a separate copy of the state in question and by
pooling results obtained from such measurement we want to distinguish
different states.

Pure states are, in fact, points in the projective space
$\mathbb{P}(\mathcal{H})$ rather than vectors in $\mathcal{H}$ -- two vectors
$\ket{v},\ket{w}\in\mathcal{H}$ represent the same state if and only if
$\ket{v}=z\ket{w}$, where $z\in\mathbb{C}$. Identification of all vectors
differing by a multiplicative complex constant gives a point in
$\mathbb{P}(\mathcal{H})$. We will denote by $\pi(\ket{v})$ the point in the
projective space corresponding to $\ket{v}\in\mathcal{H}$, i.e.\ $\pi$
denotes the canonical projection from $\mathcal{H}$ to
$\mathbb{P}(\mathcal{H})$. The action of the unitary group on $\mathcal{H}$
as in (\ref{LUequiv}) translates \textit{via} $\pi$ to an action on
$\mathbb{P}(\mathcal{H})$. We can thus reformulate the definition of
LU-equivalent states in the language of the projective space. Two pure states
$\ket{v_1}$ and $\ket{v_2}$ are locally unitary equivalent (LU-equivalent) if
and only if there exist special unitary operators $U_i\in \mathrm{SU}(N)$
such that
\begin{equation}\label{LUequiv1}
\pi(\ket{v_1})=\pi(U_1\otimes\ldots\otimes U_M \ket{v_2}).
\end{equation}
In more technical terms LU-equivalent states lie on the same orbit of
$G=\mathrm{SU}(N)\otimes\ldots\otimes\mathrm{SU}(N)$ action on
$\mathbb{P}(\mathcal{H})$ and checking the LU-equivalence can be reduced to
investigations of the structure of such an orbits. We restricted the group to
$\mathrm{SU}(N)$ since vectors differing by a phase factor are
indistinguishable on the projective space level. Furthermore, two states
represented by arbitrary (not necessary norm one) vectors $\ket{v_1}$ and
$\ket{v_2}$ are indistinguishable by local measurements if and only if they
fulfill (\ref{indisti}) with $\ket{v_i}$ substituted by
$\frac{\ket{v_i}}{\sqrt{\bk{v_i}{v_i}}}$.

The study of orbits of the group $G$ in the projective space
$\mathbb{P}(\mathcal{H})$ is facilitated by the existence of some additional
structures on $\mathbb{P}(\mathcal{H})$ and $G$ which we now briefly
describe.

The projective space $\mathbb{P}(\mathcal{H})$ is a symplectic manifold. It
means that there exists a differential two-form $\omega$ on
$\mathbb{P}(\mathcal{H})$ which is closed ($d\omega =0$) and nondegenerate.
To define $\omega$ we have to determine its action on a pair of arbitrary
vectors $A_x$, $B_x$ tangent to $\mathbb{P}(\mathcal{H})$ in an arbitrary
point $x$. Now, each tangent vector $A_x$ is a vector tangent to the curve
$t\mapsto\pi(\exp(tA)\ket{v})$ at $t=0$, where $A$ is some element of the Lie
algebra of the full unitary group $\mathrm{U}(\mathcal{H})$ [in our case by a
choice of a a basis in $\mathcal{H}_i$ it can be identified with
$\mathrm{U}(N^M)$], and $\ket{v}$ is a vector in $\mathcal{H}$ such that
$\pi(\ket{v})=x$. We have then
\begin{equation}\label{sympproj4}
\omega(A_x,B_x)=-\mathrm{Im}\frac{\bk{Av}{Bv}}{\bk{v}{v}}=\frac{i}{2}\frac{\bk{[A,B]v}{v}}{\bk{v}{v}}.
\end{equation}
It is easy to check that the above definition is correct, i.e.\ it does not
depend on the choice of $\ket{v}$ and moreover $\omega$ is invariant with
respect to the action of $U(\mathcal{H})$ on $\mathbb{P}(\mathcal{H})$, i.e.\
the action of $U(\mathcal{H})$ is \textit{symplectic}. This fact is a basis
for another canonical construction in symplectic geometry, namely the
\textit{moment map} \cite{guillemin84}. Let us denote bu
$\mathfrak{u}(\mathcal{H})$ the Lie algebra of the unitary group
$\mathrm{U}(\mathcal{H})$, i.e.\ the algebra of all anti-Hermitian operators
acting on $\mathcal{H}$ and by $\mathfrak{u}^\ast(\mathcal{H})$ the space
dual to $\mathfrak{u}(\mathcal{H})$, i.e. the space of all linear operators
on $\mathfrak{u}(\mathcal{H})$. The latter can be identified with the space
of all Hermitian operators (observables) on $\mathcal{H}$ and the action of
$Y\in\mathfrak{u}^\ast(\mathcal{H})$ on $X\in\mathfrak{u}(\mathcal{H})$ is
given by $\langle Y,X\rangle=\tr(XY)$, where in order to calculate the trace
we may choose an arbitrary basis in $\mathcal{H}$ and treat all operators as
appropriate matrices.

Instead of giving the formal definition of the moment map $\mu$ in
the most general setting (see \cite{guillemin84}) we write
explicitly an expression valid in the considered case where
$\mu:\mathbb{P}(\mathcal{H})\rightarrow
\mathfrak{u}^\ast(\mathcal{H})$ can be shown to read as
\begin{equation}\label{momproj}
\langle\mu(x),A\rangle=\frac{i}{2}\frac{\bk{v}{Av}}{\bk{v}{v}},
\quad x=\pi(\ket{v}).
\end{equation}

Besides of its fundamental mathematical origin the idea of moment map has a
natural physical interpretation. Namely to any state $\ket{v}$ we assign a
linear functional $\mu(\pi(\ket{v}))\in \mathfrak{u}^\ast(\mathcal{H})$ which
encodes information about all expectation values of all observables in the
state $\ket{v}$. In the following we will be interested in the restrictions
of $\omega$ and $\mu$ to an orbit $G.x$ of the group $G$ through a point
$x\in \mathbb{P}(\mathcal{H})$. They can be calculated as in
(\ref{sympproj4}) and (\ref{momproj}) but now $A$ and $B$ are restricted to
elements of
$\mathfrak{g}=\mathfrak{su}(N)\oplus\ldots\oplus\mathfrak{su}(N)$, i.e., the
Lie algebra of $G$. In case of the moment map it means that to every state
$\ket{v}$ we assign an element of $\mathfrak{g}^\ast$ which encodes
expectation values of Hermitian operators of the following type
\begin{equation}\label{op}
Y_1\otimes I\otimes\ldots\otimes I+\ldots+I\otimes \ldots\otimes
I\otimes Y_M.
\end{equation}
It means we extract from state $\ket{v}$ the information about local
measurements as we only control operators of (\ref{op}) type.

Observe that we can identify two natural actions of the group
$U(\mathcal{H})$. The first one is the action on $\mathbb{P}(\mathcal{H})$
induced form the ordinary action on $\mathcal{H}$ itself, whereas the second
is the \textit{coadjoint action} on $\mathfrak{u}^\ast(\mathcal{H})$ defined
as $Y\mapsto UYU^\dagger$. Both are interwined by the moment map. It can be
directly checked using (\ref{momproj}) that $\mu$ is \textit{equivariant},
\begin{equation}\label{eq}
\langle\mu(\pi(U\ket{v})),A\rangle=\langle U\mu(\pi(\ket{v})U^\dagger,
A\rangle.
\end{equation}
The moment map projects thus the orbit through $x=\pi(\ket{v})$ on the orbit
of the coadjoint action through $\mu(\pi(\ket{v}))$. As already mentioned we
may restrict the reasoning to the actions of $G$ which is a subgroup of
$U(\mathcal{H})$. Generally the mapping by the moment map will not be a
diffeomorphism between the orbit $G.x$ and the coadjoint orbit through
$\mu(\pi(\ket{v}))$. The latter operator can be stabilized by some subgroup
$Stab(\mu(\pi(\ket{v})))$ of $G$, i.e.,
$U\mu(\pi(\ket{v}))U^\dagger=\mu(\pi(\ket{v}))$ for some nontrivial $U$.
Likewise, we may have  $\pi(V\ket{v})=\pi(\ket{v})$ for some nontrivial
unitaries $V$ forming a subgroup $Stab(\pi(\ket{v}))$ of $G$. In general the
stabilizers $Stab(\pi(\ket{v}))$ and $Stab(\mu(\pi(\ket{v})))$ are not equal,
so the mapping is not diffeomorphic. The other facet of this fact is that the
form $\omega$ restricted to $G.x$ is no longer nondegenarate; $G.x$ is not a
symplectic space. The difference between dimensions of both stabilizer
subgroups conveys an important information useful in deciding the
LU-equivalence as will be shown in the second example further in the text.

The fact that $\mu$ maps orbits of $G$ in $\mathcal{H}$ on orbits of
the coadjoint action is crucial for identifying states on the same
orbit. The relevant observation is that each coadjoint orbit
intersects the subspace in $\mathfrak{g}^*$ which is dual to the
maximal commutative subalgebra of $\mathfrak{g}$ \cite{guillemin84}.
In our case this subspace consists of operators of the form
(\ref{op}) with $Y_i$ diagonal. Let us explain how this fact is
reflected in the present setting. To this end we consider a state
\begin{equation}\label{psi}
\ket{v}=\sum_{i_1,\ldots,i_M=1}^N C_{i_1\ldots
i_M}\ket{i_1}\otimes\ldots\otimes \ket{i_M}.
\end{equation}
>From the coefficients $C_{i_1\ldots i_M}$ we can build $M$ following
$N\times N$ Hermitian positive semidefinite matrices
\begin{equation}\label{psiprime}
(C_k)_{\hat{i_k}\hat{j_k}}=\bar{C}_{i_1\ldots \hat{i_k}\ldots
i_M}C_{i_1\ldots \hat{j_k}\ldots
i_M},
\end{equation}
where by the overbar we denoted the complex conjugation and the summation
over repeating indices is assumed. The orbit of $G$ through $x=\pi(\ket{v})$
is now mapped by $\mu$ on some coadjoint orbit and the fact that the latter
contains a point (\ref{op}) with $Y_i$ diagonal means, after translating back
to the orbit $G.x$, that it contains a point $x^\prime=\pi(\ket{v^\prime}$,
$\ket{v^\prime}=\sum_{i_1,\ldots,i_M=1}^N C^\prime_{i_1\ldots
i_M}\ket{i_1}\otimes\ldots\otimes \ket{i_M}$ for which the corresponding
matrices $C_k^\prime$ are diagonal,
$C_k^\prime=\mathrm{diag}(p_{1k}^2,\ldots,p_{Nk}^2)$. The numbers $p_{lk}^2$
have a straightforward interpretation as probabilities to find the $k$-th
subsystem in the state $\ket{l}$ while neglecting other $M-1$ subsystems (see
\cite{shk10} for detailed calculations). Let us remark that in the case of
two subsystems, $M=2$, the whole procedure leads to the familiar Schmidt
decomposition which allows for a straightforward identification of
LU-equivalent states for bipartite systems. There is no useful generalization
of the Schmidt decomposition for more than two subsystems, nevertheless, the
following reasoning shows how to overcome this obstacle.

To pass from $\ket{v}$ to $\ket{v^\prime}$ one has to act on
$\ket{v}$ by $\tilde{U}_1^T\otimes\ldots\otimes \tilde{U}_M^T$ ($^T$
means the transposition), where $\tilde{U}_k\in \mathrm{SU}(N)$ are
such that $\tilde{U}_k^\dagger
C_k\tilde{U}_k=\mathrm{diag}(p_{1k}^2,\ldots,p_{Nk}^2)$. It is
always possible to find such unitaries since, as already mentioned,
$C_k$ are Hermitian. Moreover one  can always choose $\tilde{U}_k$
in a such a way that $p_{1k}^2\geq\ldots\geq p_{Nk}^2$ and in the
following we assume it was done. In fact all intersections of the
coadjoint orbit with (\ref{op}) with diagonal $Y_i$ differ only by
ordering of the diagonal elements. We avoid thus this ambiguity by
fixing the order.

The image of $\pi(\ket{v^\prime})$ under the action of the moment
map $\mu$ can be calculated \cite{shk10}. It has the form (\ref{op})
with
\begin{equation}\label{cart}
Y_k=\alpha\mathrm{diag}(-\frac{1}{N}+p_{1k}^2,\ldots,-\frac{1}{N}+p_{Nk}^2),
\end{equation}
where $\alpha$ is some appropriate const. Going back to the above mentioned
physical interpretation of $\mu$ we see that in fact moment map encodes
information about the state $\ket{v^\prime}$ contained in local measurements
as it is determined only by the local probabilities $p_{kl}^2$.

The matrices $C_k^\prime$ and, consequently $Y_k$ have, in general,
degenerate spectra, i.e, several $p_{lk}$ can repeat in
(\ref{cart}). Let us denote by $\nu_k$ the number of different
eigenvalues of $C_k^\prime$ and by $m_{k,n}$ the multiplicity of the
$n$-th eigenvalue. It is now easy to show \cite{shk10} that the
stabilizer $Stab(\mu(\pi(\ket{v^\prime})))$ consists of
$U_1\otimes\ldots\otimes U_M \in G$ where
\begin{equation}\label{nu}
U_k=\left(
          \begin{array}{cccc}
            u_{k,0} &  &  &  \\
             & u_{k,1} &  &  \\
             &  & \ddots & \\
             &  &  & u_{k,\nu_k} \\
          \end{array}
        \right),
\end{equation}
with $u_{k,n}\in \mathrm{U}(m_{k,n})$. The equivariance (\ref{eq}) of $\mu$
implies that  $Stab(\pi(\ket{v^\prime}))\subset
Stab(\mu(\pi(\ket{v^\prime})))$. Consequently, for any $\ket{v}$ the
corresponding state $\ket{v^\prime}$ is given up to the action of
$Stab(\mu(\pi(\ket{v^\prime})))$, which is explicitly known.

We thus arrive at the following algorithm to check if two states $\ket{v_1}$
and $\ket{v_2}$ are LU-equivalent.
\begin{itemize}
  \item For both states compute the matrices $C_k$ (\ref{psiprime}) and
      check whether they have pairwise the same ordered spectra. If this
      is not the case the states are not LU-equivalent.
  \item If the first condition is fulfilled use the unitary matrices
      $\tilde{U}_k$ diagonalizing $C_k$ for each state to find
      $\ket{v_1^\prime}$ and $\ket{v_2^\prime}$. If they are equal then
      $\ket{v_1}$ and  $\ket{v_2}$ are LU-equivalent, otherwise they are
      LU-equivalent if and only if there exist $U_1\otimes\ldots\otimes
      U_M \in Stab(\mu(\pi(\ket{v_i^\prime})))$ such that
      $\ket{v_1}=U_1\otimes\ldots\otimes U_M \ket{v_2}$.
\end{itemize}
In case of generic states $Stab(\mu(\pi(\ket{v_i^\prime})))$ consists of
diagonal matrices $\mathrm{diag}(e^{i\phi_1},\ldots,e^{i\phi_N})$ and above
method can be easily applied. For other states our method also simplifies
computations since we no longer need to use the full $U(N)\otimes \ldots
\otimes U(N)$ group but can restrict our attention to an appropriate
subgroup. The worst case is when the state $\ket{v}$ is such that all $C_k$
are $\frac{1}{\sqrt{N}}\mathbb{I}$. Then the action of $U(N)\otimes \ldots
\otimes U(N)$ on such state gives a state for which again all $C_k$ are
normalized identity matrices.

Our approach gives also a beautiful mathematical characterization of states
which are indistinguishable by local measurements. As we noticed before for
any state $\ket{v}$ the moment map $\mu$ assigns an element of
$\mathfrak{g}^\ast$ which encodes all expectation values of local
measurements in state $\ket{v}$. In other words all states that have the same
expectation values for local measurements are sent by the moment map to the
same point of $\mathfrak{g}^\ast$ -- they constitute a fiber over this point.
To examine if states $\ket{v_1^\prime}$ and $\ket{v_2^\prime}$ obtained in
the second step of the above algorithm are indistinguishable by local
measurements it is enough to check if ordered spectra of corresponding
matrices $C_k$ are the same. An intimate connection between LU-equivalence
and indistinguishability by local measurements appears to be helpful in
deciding the former in various situations, also in the `worst case' mentioned
above, as we show in the second example below.

A detailed analysis showing how to apply our method to different kind of
states will be published elsewhere \cite{sk10u}. Here we give only two simple
examples. The first one shows how the proposed algorithm works in a generic
case and involves three qutrits. The Hilbert space is
$\mathcal{H}=\mathbb{C}^{3}\otimes\mathbb{C}^{3}\otimes\mathbb{C}^{3}$ and
has the real dimension $\mathrm{dim}_{\mathbb{R}}(\mathcal{H})=54$. The
dimension of the projective space $\mathbb{P}(\mathcal{H})$ is thus
$\mathrm{dim}_{\mathbb{R}}(\mathbb{P}(\mathcal{H}))=52$. We are interested in
the orbits of $G=SU(3)\otimes SU(3)\otimes SU(3)$ - action on
$\mathbb{P}(\mathcal{H})$. The Lie algebra $\mathfrak{g}$ of $G$ is spanned
by $A\otimes I\otimes I$, $I\otimes A\otimes I$, and $I\otimes I\otimes A$
with $A\in \mathfrak{su}(3)$ and
$\mathrm{dim}_{\mathbb{R}}(\mathfrak{g})=24$. Let us consider two states
\begin{gather*}
\ket{\Psi}=\alpha\ket
W+\beta\ket{222}=\alpha(\ket{001}+\ket{010}+\ket{100})+\beta\ket{222},\\\nonumber
\ket{\Phi}=\sqrt{2}\alpha \ket{000}+\alpha\ket{111}+\beta\ket{222},
\end{gather*}
 where $3|\alpha|^{2}+|\beta|^{2}=1$ and $|\beta|<|\alpha|$. Direct
 computations show that for both $\ket{\Psi}$ and $\ket{\Phi}$ we get:
\begin{gather}\label{three_qut_ex}
C_{1}=C_2=C_3=\left(\begin{array}{ccc}
2|\alpha|^{2} & 0 & 0\\
0 & |\alpha|^{2} & 0\\
0 & 0 & |\beta|^{2}\end{array}\right).
\end{gather}
Since the corresponding matrices for both states are the same,
$\ket{\Phi}$ and $\ket{\Psi}$ are indistinguishable by local
measurements. Remember that if for at least one pair of the matrices
$C_i$ calculated for $\ket{\Phi}$ and $\ket{\Psi}$ the results
differ the algorithm ends deciding that the states are not
LU-equivalent. Since this is not the case we have to invoke the
second step. The matrices (\ref{three_qut_ex}) are already diagonal
so we do not need to change the form of $\ket{\Psi}$ and
$\ket{\Phi}$. Let us thus look at the states that are $LU$ -
equivalent with $\ket{\Phi}$ and for which matrices $C_1,C_2,C_3$
are given by (\ref{three_qut_ex}). Such states are generated by
action of $\mathrm{Stab}(\mu(\ket{\Phi}))$ on $\ket{\Phi}$. But
$\mathrm{Stab}(\mu(\ket{\Phi}))$ consists of matrices $U_1\otimes
U_2 \otimes U_3$ where:
\begin{equation}\label{stabi}
U_k=\left(
          \begin{array}{ccc}
            e^{i\phi_{k1}}& 0 & 0   \\
             0 & e^{i\phi_{k2}} & 0   \\
             0 & 0 & e^{-i(\phi_{k1}+\phi_{k2})} \\
          \end{array}
        \right),
\end{equation}
The action of (\ref{stabi}) on $\ket{\Phi}$ gives
\begin{equation}\label{psilu}
\ket{\Phi^\prime}=\sqrt{2}\alpha e^{i\phi_1}\ket{000}+e^{i\phi_2
}\alpha\ket{111}+e^{i\phi_3}\beta\ket{222}
\end{equation}
where $\phi_i$ are appropriate combinations of $\phi_{ki}$. It is easy to
notice that $\ket{\Psi}$ is not of the (\ref{psilu}) form and in the effect
states $\ket{\Psi}$ and $\ket{\Phi}$ are not $LU$ - equivalent.

Fibers of the moment map consisting, as mentioned above, of locally
indistinguishable states are given as common level sets $\{f_A(x)=c_A\}$ of
functions $f_{A}(\pi(v))=\langle\mu(\ket{v}),A\rangle$ [see (\ref{momproj})],
where $A\in\mathfrak{g}$ and $c_A$ are some constants. An interesting
question which one should ask is what is the relationship between the tangent
space to the fiber of the moment map at given point $x=\pi(\ket{v})$ and the
tangent space to the orbit of $G$-action at the same point. These two tangent
spaces generate infinitesimal movements in $\mathbb{P}(\mathcal{H})$ in the
direction of states which are indistinguishable by local measurements from
state $x$ and, respectively, $LU$-equivalent with $x$. The tangent space to
the fiber at a given point $x=\pi(\ket{v})$ is of course contained in the
intersection of kernels of all $1$-forms $d_xf_A$, where $A$ is an element of
$\mathfrak{g}$. It is very important to emphasize that it might be just a
subspace of this intersection \cite{sk10u}.

To continue our discussion we will need the following definition.
\begin{defi}
Let $(V,\omega)$ be symplectic vector space and $W\subset V$ its subspace.
Then $\omega$-orthogonal complement of $W$ is given by
\begin{equation}\label{omegaorth}
W^{\bot\omega}=\{v\in V:\,\,\omega(w,v)=0\,\,,\forall w\in W\}.
\end{equation}
\end{defi}
In \cite{sk10u} we prove that the tangent space to the orbit of
$G$-action at $x$ is spanned by vectors which are
$\omega$-orthogonal to the intersection of kernels of all $1$-forms
$d_xf_A$ for $A\in\mathfrak{g}$. Knowing just these kernels we can
thus find the dimension of an orbit of the $G$-action. The other
facet of the above reasoning is that when the tangent space to a
fiber of the moment map is contained in the tangent space to the
orbit $G.x$ then the local unitary equivalence is the same as the
indistinguishability by local measurements, i.e., in this case we
can restrict our method just to the comparison of ordered spectra of
$C_k$ matrices. In \cite{shk10} we prove that the dimension of the
part of the moment map which is contained in the orbit of the
$G$-action is equal to difference between dimension of stabilizers
of $\pi(\ket{v})$ and $\mu(\pi(\ket{v}))$ mentioned above.

The next example shows a practical application of the above observation. Let
us consider a three-qubit state
\begin{equation}\label{GHZ}
\ket{GHZ}=\frac{1}{\sqrt{2}}(\ket{000}+\ket{111})
\end{equation}
Here $\mathrm{dim}_{\mathbb{R}}(\mathbb{P}(\mathcal{H}))=14$. One calculates
$C_{1}=C_2=C_3=\frac{1}{2}\mathbf{1}_{2\times 2}$. Hence
$\mathrm{Stab}(\mu(\ket{GHZ})$ is the whole local unitary group, i.e,
$Stab(\mu(\ket{GHZ})=SU(2)\otimes SU(2)\otimes SU(2)$,
and its action on $\ket{GHZ}$ is the whole orbit $\mathcal{O}_{\ket{GHZ}}$.
Thus the part of the fiber of the moment map contained in the orbit
$\mathcal{O}_{\ket{GHZ}}$ is exactly the orbit $\mathcal{O}_{\ket{GHZ}}$. The
intersection, $K$, of kernels of $df_{A\otimes I\otimes I},\, df_{I\otimes
A\otimes I},\,df_{I\otimes I\otimes A}$, where $A\in \mathfrak{su}(3)$ is
seven dimensional. The dimension of the $\omega$-orthogonal complement of $K$
is the dimension of the $G$-orbit through $\ket{GHZ}$ and is equal
$\mathrm{dim}(\mathcal{O}_{\ket{GHZ}})=7$.
The maximal possible dimension of the fiber is also $7$ since it is the
dimension of $K$. This means that the whole fiber of the moment map is
contained in $\mathcal{O}_{\ket{GHZ}}$. We infer thus that two three-qubit
states for which $C_{1}=C_2=C_3=\frac{1}{2}\mathbf{1}_{2\times 2}$ are
automatically $LU$-equivalent, although the situation corresponds the `worst
case' when deciding the LU-equivalence should be the most difficult.

The presented examples do not exhaust the variety of possible situations met
when checking LU-equivalence \cite{sk10u}. They correspond to, in a sense,
two extremal cases, and show clearly that dimensionality of subsystems does
not play a crucial role in applications, although, admittedly can make
calculations more cumbersome.

The support by SFB/TR12 'Symmetries and Universality in Mesoscopic Systems'
program of the Deutsche Forschungsgemeischaft and Polish MNiSW grant no.\
DFG-SFB/38/2007 is gratefully acknowledged.

\bibliographystyle{apsrev}

\end{document}